\documentclass{article}
\usepackage[english]{babel}

\usepackage[a4paper,top=2cm,bottom=2cm,left=2cm,right=2cm,marginparwidth=1.75cm]{geometry}

\usepackage{amsmath}
\usepackage{graphicx}
\usepackage[]{hyperref}
\usepackage{titlesec}
\usepackage{upgreek}
\usepackage{lipsum}
\usepackage[doublespacing]{setspace}
\usepackage[normalem]{ulem}
\usepackage{textcomp}
\usepackage{eurosym}
\usepackage{authblk}
\titleformat{\section}{\centering\bfseries\normalsize}{\thesection}{1em}{}
\titleformat{\subsection}{\slshape}{\thesubsection.}{2.65em}{}
\titleformat{\subsubsection}{\slshape}{\thesubsubsection.}{1.8em}{}

\title{\textbf{Using low-cost Blu-Ray Optical Pickup Units for Measurement of Single Photon Emission from NV-Centers}}

\author[1]{\normalsize Simon Klug}
\author[1]{Jonas Homrighausen}
\author[2]{Peter Glösekötter}
\author[3]{Andreas W. Schell}
\author[1]{Markus Gregor}

\affil[1]{\textit{Department of Physics Engineering, FH Muenster - University of Applied Sciences, Stegerwaldstrasse 39, D-48565 Steinfurt, Germany}}
\affil[2]{\textit{Department of Electrical Engineering and Computer Science, FH Muenster - University of Applied Sciences, Stegerwaldstrasse 39, D-48565 Steinfurt, Germany}}
\affil[3]{\textit{Institute of Semiconductor and Solid State Physics, Johannes Kepler University Linz, Altenberger Str 69, 4040 Linz, Austria}}

\date{} 

\begin{document}

\maketitle

\begin{abstract} 
This work presents a cost-effective method for collecting single photons emitted from single nitrogen-vacancy centers in nanodiamonds. Conventional components of a confocal laser-scanning microscope, such as microscope objectives and the piezo translation stages, are replaced by two affordable Blu-ray optical pickup units. A Hanbury Brown and Twiss setup is used to identify single photon emission. The proposed approach is inexpensive and simple and lowers the entry-level to single photon research for quantum technologies. This enables student lab experiments or demonstration experiments at schools and shows that efficient sources of quantum light can be made from standard components compatible with established industry processes.

\end{abstract}

\section{Introduction}
In the developing field of quantum technologies, single photon sources (SPSs) have attracted considerable interest due to their role in various quantum-based applications, including quantum communications, cryptography, computing, and metrology \cite{childress2013, Pezzagna2021,Rondin2014, Balasubramanian2008, Maze2008}. These applications rely significantly on the controlled generation and coupling of single photons. Among the various methods investigated for single photon generation, the use of single nitrogen-vacancy centers (NV-centers) in diamonds has proven to be one of the promising and versatile solutions, since they exhibit efficient photon emission at room temperature, photostability, and offer easy ways to manipulate its spin quantum state \cite{Dutt2007, Doherty2013}.

There are different approaches for interfacing single photons, such as using high-NA microscope objectives with piezo positioning systems for accurate sub-micrometer positioning \cite{Kurtsiefer2000, Rodiek2017, Beveratos2002, Kim2018}, direct coupling to fibers \cite{Schroeder2010, Schroeder2012, Liebermeister2014} or employing special optical elements \cite{Schell2014, Preuss2022, Schwab2022}, although these methods are usually expensive to produce or build.
The goal of this work was to develop a cost-effective solution based on a Confocal Laser Scanning Microscope (CLSM) setup for the efficient optical coupling of single quantum emitters based on NV centers. A major challenge is to identify an alternative option for precise and selective control of individual diamonds without relying on expensive piezoelectric technology and replacing expensive optical components like high-NA microscope objectives. 
 In this work, a commercially available optical pickup unit (OPU) (Blu-Ray, PHR-803T, Toshiba Co., Tokyo, Japan), which has already been successfully used in other low-cost applications \cite{Hwu2018, Gaudi, diyouware}, is used to replace the high-NA microscope objective and piezoelectric positioner. 

\section{Experimental Setups}

\begin{figure}[ht!]
\centering\includegraphics[width=0.5\linewidth]{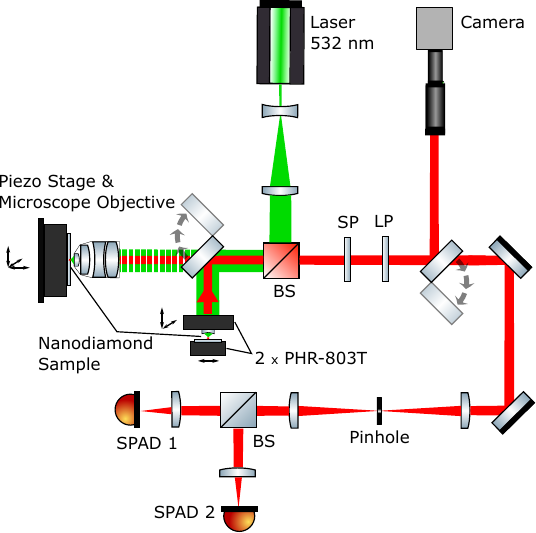}
\caption{Schematic drawing of the CLSM setup including the piezo translation stage and High-NA microscope objective and, also, the PHR-803T Blu-Ray pickups as translation stages and lens. A flip mirror can be used to switch between both setups.}
\label{fig:clsm}
\end{figure}

\subsection{Reference CLSM Setup}

CLSMs have proven to be a powerful technique for detecting individual NV centers making it an ideal choice for investigating individual nanodiamonds on a coverslip. 

To get more detailed information about the diamond samples and to get a comparison for a potential 2D scan with the low-cost setup, a reference CLSM is built (see Fig. \ref{fig:clsm}). Therefore, a single frequency diode-pumped laser (Cobolt 08-DPL) with a good beam quality ($M^2 < 1.1$) is expanded by a Galilean telescope (Thorlabs ACN127-030-A + Thorlabs LA1708-AB) with a magnification of  $M=6.\bar{6}$  from approx. $\approx 0.6 \:\textrm{mm}$ to approx. $\approx 4 \:\textrm{mm}$ in diameter to fully illuminate the lens system inside the OPU. The beam is reflected by a dichroic beamsplitter (Thorlabs DMLP550R) and focused on the sample through a microscope objective (Nikon CFI Plan Apo Lambda 60XC) with a high NA of 0.95. To eliminate the laser- and background light, the collected fluorescence is spectrally filtered by the dichroic beam splitter, two long-pass (Thorlabs FELH0550) and one short-pass (Thorlabs FESH0750) filter. Following this, a spatial frequency filter ($30  \: \upmu\textrm{m} \:  \textrm{pinhole}, \:2 \: \times \: \textrm{Edmund Optics} \: 48774 \: \textrm{f=100 mm}$) is used to eliminate out-of-focus light. To perform a 2d confocal scan of a sample, a piezo translation stage (PiezoSystem Jena Tritor 100, open loop) is used to allow movements with nanometer precision and ensure accurate positioning. For the Hanbury Brown and Twiss (HBT) experiment, two actively quenched and cooled SPAD (Laser Components COUNT T-series) and the correlation electronics (TimeTagger 20, Swabian Instruments) are utilized. Furthermore, a scientific CMOS camera (Hamamatsu Orca Fusion BT) for widefield microscopy for pre-aligning the nanodiamond sample slide is integrated into the beam path.

\subsection{Low Cost Single Photon Coupling}

NV-centers can exist in the NV$^-$ negative or NV$^0$ neutral charge state and both can act as a source of single photons. The NV-centers can efficiently be excited using a laser at a wavelength of $532 \:\textrm{nm}$. The fluorescence spectrum for an ensemble of NC-centers inside a $100 \: \upmu\textrm{m} $ diamond is shown in figure  \ref{fig:spectrum} (a). Both charge states can be identified in the sample by the zero phonon line at $575 \:\textrm{nm}$ \cite{Zhang2024} for the NV$^0$-center and $638\:\textrm{nm}$  \cite{Schirhagl2014} for the NV$^-$-center. At room temperature, the emission is broader ranging from $630  \:\textrm{nm}$ to $800 \:\textrm{nm}$. The lifetimes of the two charge states in nanodiamonds differ slightly. The NV$^-$ centers has a lifetime of $12\:\textrm{ns}$ compared to $21 \:\textrm{ns}$ for the NV$^0$ center \cite{Storteboom2015}. This makes the NV$^-$ center brighter and more easily to be detected. A spectrum of a bulk diamond with $1\:\upmu\textrm{m}$ in diameter containing NV centers in both charge states was captured using a spectrometer (Ocean HDX spectrometer) in the described optical setup and is exemplarily depicted in Figure \ref{fig:spectrum} (a) (red).

\begin{figure}[t!]
\centering\includegraphics[width=0.45\linewidth]{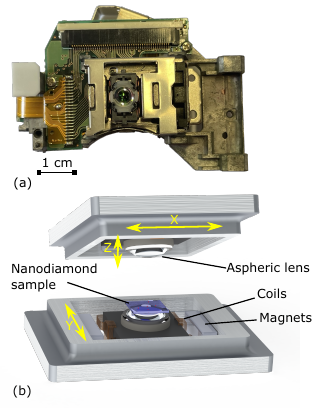}
\caption{(a) Single PHR-803T Blu Ray optical pickup unit (Toshiba Co., Tokyo, Japan). (b) Configuration of two pickups rotated by 90° to perform a confocal 2D scan. The upper pickup is used as a microscope objective while the sample is glued to the lower one. }
    \label{fig:laufwerk}
\end{figure}

As a low-cost approach, and as a replacement for the microscope objective to collect the photons, the integrated lens from a PHR-803T Blu-Ray pickup is to be used (see Fig. \ref{fig:laufwerk} (a)). As shown in Figure \ref{fig:clsm}, it is integrated into the existing setup using a flip mirror. The aspheric lens has a large NA of about 0.6 at 650 nm \cite{bluray}. Transmission measurements for this optic were taken using a spectrophotometer (PerkinElmer Lambda 1050), which shows that the lens has an anti-reflective coating and high transmission, particularly in the spectral range of the NV center emission (see Fig. \ref{fig:spectrum} (a) (black)). The OPU has a smaller transmission at the pump laser wavelength and can be compensated by increased laser power.

\subsection{Scanning OPU Setup}

To be able to perform spatial scans without employing piezo actuators, the internal positioning mechanisms of the PHR-803T are utilized for the XY-scan. The optics of a single OPU can be moved along the horizontal and vertical direction and can also be tilted in the horizontal direction. The operation principle is similar to voice-coil actuators.
Five individual coils are directly fixed to the lens holder.  Two permanent magnets near the coils create an external static magnetic field and allow the optics to be positioned precisely. For the laboratory demonstrator, the OPUs were clamped in optical mounts and placed opposite each other. In order to realize the 2D scan, it is necessary to install two PHR-803T at an angle of $90^ \circ$ to each other, as seen in Figure \ref{fig:laufwerk} (b) \cite{Gaudi}. A solid 3D-printed housing is intended for future setups.

\begin{figure}[t!]
\centering\includegraphics[width=0.6\linewidth]{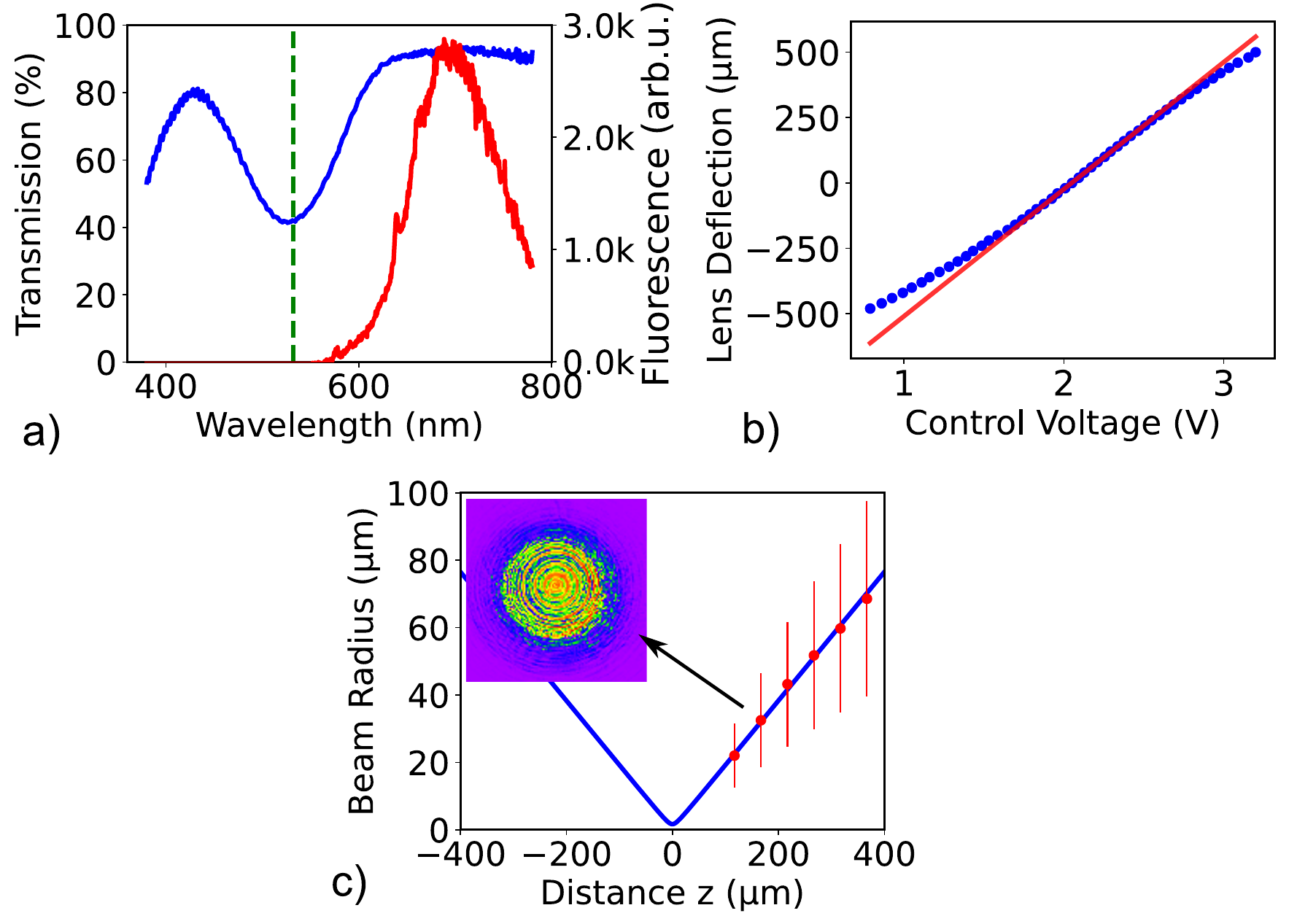}
\caption{(a) Transmission spectrum of the PHR-803T lens (blue), excitation laser wavelength (green), and emission spectrum of NV center (red). (b) Bidirectional deflection of the lens for the horizontal axis (blue) and linear fit (red) to illustrate the deviation. (c) Measurement of the beam radii at several positions (red) and fit of the transverse Gaussian profile (blue) to determine the excitation beam radius $w_0$. Inset shows the mode profile $167 \:\upmu\textrm{m} $ behind the OPU aspheric lens.} 
    \label{fig:spectrum}
\end{figure}

Since the lens is mounted in the center position and is held by four springs, it can deflect in two directions.  Therefore, the direction of the currents within coils needs to be controlled. To ensure this, DC-motor control operational amplifiers (TCA0372) were used \cite{Gaudi}, which provide an output current of up to one ampere. The input voltages for the scan are provided by the 16-bit data digital-to-analog converter (DAC, RedLab 3103 USB), which was also used to provide the input voltages for the piezo driver in the reference setup. In the DC motor circuit used here, the virtual ground is at $\approx 2 \:\textrm{V}$ output voltage of the DAC. Here, the lens is in the center position and can be deflected bidirectionally with a control voltage between $1 \:\textrm{to}\: 3 \:\textrm{V}$, resulting in coil voltages between -$1 \:\textrm{to} \:1 \:\textrm{V}$ and coil currents between -$200 \:\textrm{to}\: 200 \:\textrm{mA}$.  The deflection in the horizontal direction in  Figure \ref{fig:spectrum} (b) was measured by continuously increasing the current through the coils and tracking it with a manual translation stage. The deflection has a high linearity in both directions around the center position of the lens.

One of the factors limiting the resolution of a CSLM is the size of the exciting laser spot on the sample. For the nanodiamonds to be excited individually, the distribution on the slide and thus the minimum distance between the individual diamonds needs to agree with the diameter of the laser spot. Since these OPUs are not optimized for the green wavelength of the excitation laser and no information about the focal length is available, an optical approach was chosen to determine the focus size. We positioned a CMOS image sensor (Raspberry Pi High Quality Camera Module) at various distances behind the lens. Gaussian fits through the center of the mode images (an example is shown in Figure \ref{fig:spectrum} (c)) were used to measure diameters at different positions. Subsequently, the transverse Gaussian beam profile $w(z) = w_0 \sqrt{1 + \left( \frac{z \lambda}{\pi w_0^2} \right)^2}$ was fitted for all measured diameters, as shown in Figure \ref{fig:spectrum} (c). From this fit, a beam radius in the focus of $ w_0 = 1.66 \pm 0.74 \:\upmu\textrm{m} $ can be estimated. The parameters for spin coating should therefore be selected so that the diamonds are at least $2 \cdot w_0$ apart. The mode profile pictured in Figure \ref{fig:spectrum} (c) appears suboptimal, suggesting that the smallest achievable focus diameter may not be reached, possibly due to aberration in the OPUs imaging system, since it is not optimized for our external pump wavelength of $532\:\textrm{nm}$. Nevertheless, the estimated spot diameter is sufficient for our intended measurements.

\subsection{Samples Preparation}

 Fused silica serves as an ideal coverslip substrate for nanodiamond adherence due to its minimal autofluorescence. This reduces unwanted background signals and ensures a better signal-to-noise ratio, which is very important considering the small fluorescence signal of single NV centers.

To achieve a uniform deposition of nanodiamonds (Adamas Nanotechnologies NDNV40nmLw10ml) on the coverslip, a multi-step process is employed. A concentration of 1:100 nanodiamonds in deionized water (DI-water) is explored, with a solution comprising 1 mg/ml diamonds in DI-water and $0.3 \:\textrm{wt}\%$ PVA to enhance adhesion and distribution. The coverslips undergo pretreatment with a 2\% Hellmanex solution in DI-water, as suggested in \cite{Hirt2021}, utilizing ultrasonic bath exposure at $50^\circ C$ for 10 minutes to enhance cleaning. This step minimizes contamination from residues, oils, and particles while improving nanodiamond adhesion. The ultrasonic bath also aids in dispersing nanodiamonds in DI-water and PVA, ensuring a homogeneous particle distribution. Subsequently, a spin-coating technique is employed, where $30 \:\upmu\textrm{l}$ of nanodiamond suspension is placed at the coverslip center and rotated at $3500 \:\textrm{rpm}$. This process utilizes centrifugal force to create a thin, uniform nanodiamond film, offering precise control over coating density through spin speed and duration adjustments.

\section{Confocal Scan and Measurements of the NV Center}
\begin{figure}[ht!]
\centering\includegraphics[width=0.6\linewidth]{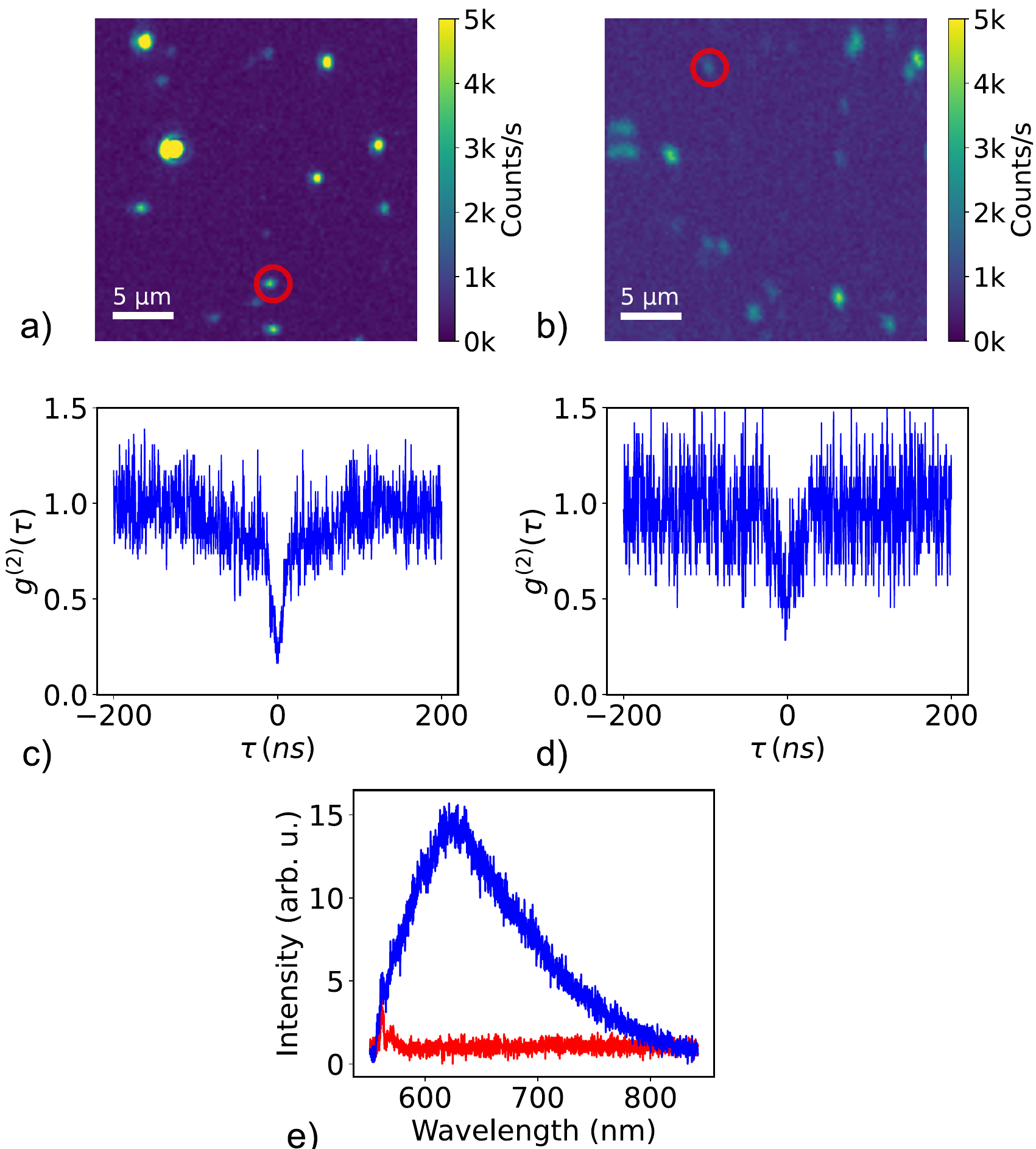}
\caption{(a) 2D confocal scan ($20 \:\upmu\textrm{m} \times 20 \:\upmu\textrm{m}$) with the reference setup to verify the homogeneous distribution of nanodiamonds. (b) 2D confocal scan ($20 \:\upmu\textrm{m} \times 20 \:\upmu\textrm{m}$) using two Blu-Ray pickups with excitation power of $P_L=10\:\textrm{mW}$, integration time of $ t_{int} = 40\:\textrm{ms}$ and resolution of $100\:\textrm{px} \times 100\:\textrm{px}$.   The red framed diamonds show clear antibunching with (c)  $g^{(2)}(0) = 0.14$ for the reference setup and (d) $g^{(2)}(0) = 0.42$ for the low-cost setup. The measurement was performed with 1000 bins, a bin width of 200 ps, and a measuring time of approx. 40 minutes. (e) Autofluorescence spectrum (Andor Kymera 328i Spectrograph) of the reference microscope objective (red) and PHR-803T aspheric lens (blue). }
    \label{fig:zeichnungclean}
\end{figure}

The 2D scanning, focusing, and data acquisition are controlled for both setups using a Python script.
The fluorescence intensity plots generated from the 2D scans indicate the quality of the control of the pickup coils. The bright spots in the image of the low-cost system (see Fig. \ref{fig:zeichnungclean}(b)), are slightly oval with a lower signal-to-noise ratio compared to those in the reference setup (see Fig. \ref{fig:zeichnungclean} (a)). This may imply that the wire resistance of the pickup coils varies slightly from coil to coil, resulting in differences in the strength of linear displacement with the control voltages.  The reduced contrast observed in the scan obtained from the low-cost setup compared to that obtained from the reference setup can be attributed to the increased autofluorescence emanating at 532 nm excitation wavelength from the polymeric aspheric lens of the OPU (see Fig. \ref{fig:zeichnungclean} (e)). Unfortunately, the autofluorescence occurs in the same spectral region as the emission spectrum of the NV center, thus reducing the signal-to-noise ratio.

The detection of single photons with low-cost optics can be demonstrated by measuring the anti-bunching behavior and comparing it with the reference setup.
For the red-marked diamonds of the reference and low-cost scans (see Fig. \ref{fig:zeichnungclean} (a), (b)), the $g^2(\tau)$-functions were measured. Surprisingly, during the measurement, the PHR-803T coils showed significantly less drift than the open-loop piezo stages of the reference setup.  An antibunching effect can be seen with both setups (see Fig. \ref{fig:zeichnungclean} (c), (d)). 
The fact that the $g^{(2)}(0)$ function reaches a value below 0.5 is a strong indicator of the presence of a single NV center.

\section{Conclusion}
To summarise, the development of mechanically stable 3D confocal scanning for measuring single photon emission is a significant advance in terms of cost efficiency and applicability in this field. By using affordable components such as two OPUs,  that cost about 30 \euro \: in sum, this approach stands in strong contrast to conventional setups, like our reference setup presented here, with expensive high-NA microscope objectives and piezo stages costing routinely more than 10k \euro.  
The development of a low-cost single photon collection system based on OPUs holds the potential to advance quantum technologies and promote their widespread adoption in industrial applications. This would also benefit schools and universities, for which quantum experiments are often too expensive and complex. The integrated lenses and voice coil actuators for positioning ensure a compact and alignment-friendly design for further experiments in single photon source designs and laser-scanning confocal microscopes.


\section*{Disclosures} The authors declare no conflicts of interest.

\section*{Data availability} The data that support the findings of this study are available from the corresponding author upon reasonable request.




\end{document}